\documentclass{aa501}
\usepackage{graphicx}
\begin{document}

\title{HD~12098 - A new northern hemisphere  roAp star.}
\author{V. Girish \inst{1,}\thanks{{\it JAP,} Dept. of physics, 
				   IISc, Bangalore-560 012}
        \and
        S. Seetha \inst{1}
        \and
        P. Martinez \inst{2}
        \and
        S. Joshi \inst{3}
        \and
        B. N. Ashoka \inst{1} 
         \and
     \\   D.W. Kurtz \inst{4,5}
         \and
        U.S. Chaubey \inst{3}
         \and
        S.K. Gupta \inst{3}
         \and
        Ram Sagar \inst{3}
       }
\institute{ISRO Satellite Centre, Airport road, Bangalore-560 017, India
        \and
        South African Astronomical Observatory, PO Box 9, Observatory 7935, South Africa
       \and
	State Observatory, Manora Peak, Naini~Tal-263 129, India
\and
	Dept. of Astronomy, University of Cape Town, Rondebosch 7700, South Africa 
\and
   Centre for Astrophysics, University of Central Lancashire, Preston PR1 2HE, UK
}
\offprints{S. Seetha (seetha@isac.ernet.in)}
\date{recd ...; accepted ......}

\abstract{
We present the analysis of 65 hours of high speed photometric observations of
HD~12098 taken from State Observatory, Naini~Tal and Gurushikhar Observatory,
Mt.~Abu on sixteen nights spanning from November 1999 to November 2000. HD~12098
is the first rapidly oscillating Ap star discovered from the   `Naini~Tal-Cape
survey for northern hemisphere roAp stars'.  It is the  32nd in the
complete list.   HD~12098 exhibits one predominant mode of  oscillation at
$\nu_1 = 2.1738$~mHz. The second-most significant frequency in our data is at
$\nu_2 = 2.1641$~mHz with a 1 cycle/day alias ambiguity. We argue that
$\nu_2$ is a rotational sidelobe of $\nu_1$, rather than an independent
pulsation mode.
Evidence for the presence of two other frequencies at 2.1807 and 2.3056~mHz
is also  presented. 
 }
\maketitle

\section{Introduction}

Many cool Ap stars exhibit anomalously strong Sr, Cr and Eu lines coupled with
 high  magnetic fields  frequently as strong as several kilogauss.  A few of these chemically peculiar
stars exhibit  broadband light oscillations  in the  4-16 minute period range with amplitudes
of few milli-magnitude (mmag). This subset of Ap stars is termed  as rapidly
oscillating Ap  (roAp) stars and the oscillations are  attributed to low-order
$(l<3)$, high overtone $(n\gg l)$ non-radial  pulsation  modes (see Kurtz~\cite{kurtz90a},
Martinez \& Kurtz~\cite{peter95} for reviews).  Long term variations in
amplitude of pulsations are observed in many roAp stars. This modulation can be
explained in the light of the oblique pulsator model.  In this model the
pulsation and magnetic axes are oblique with respect to the rotation axis of the
star.  The modulation in the amplitude is then caused by the orientation effect
due to the  periodic variation in the angle between the pulsation axis and the line of
sight.  This effect seen in many roAp stars is therefore extremely useful in
determining the rotation period of the star (Kurtz~\cite{kurtz82}).

In order to detect new roAp stars in the northern hemisphere, the {\it
``Naini~Tal-Cape Survey''}  was initiated in 1997 (Ashoka et al.~\cite{ashoka00}; Martinez et al.~\cite{peter01}),  because 28 of the previously
known 31 roAp stars are in the southern hemisphere.  HD~12098 is the first roAp
star discovered from this survey.  The co-ordinates of this star ($\alpha_{2000}
= 02^h\,00^m.7 \,,\,\delta_{2000} = 58^\circ\,31'.6$ ) make it the northern-most
known roAp star.  HD~12098  is of spectral type F0 (Olsen~\cite{olsen80}) and  has an apparent magnitude  $m_V=7.974$. It has  Str\"omgren color indices  $(b-y)=0.191$, $m_1=0.328,\, c_1=0.517,\,\beta=2.796$ (Hauck \& Mermilliod~\cite{hauck98}).  $\delta m_1=-0.122$ and $\delta c_1=-0.279$ are calculated using the calibrations of Crawford (\cite{crawford75}). Recently Wade et al.  (\cite{wade01}) have detected a  variable longitudinal magnetic field of $\approx$ 2kG in HD~12098. They also predict a rotational period of  the order of one week.   HD~12098 was discovered in 1999 from Naini~Tal as a roAp star having a  period of 7.68 minute with an evidence for multi-periodicity (Martinez et al.~\cite{ibvs4853}) and amplitude modulation. We obtained about 65 hours of high speed photometric data as part of follow-up observations out of which around 45 hours are from Mt.~Abu during October 2000. Analysis of these data indicates the presence of multi-periodic oscillations in the star.  Among these, one frequency is most likely  due to the rotational modulation of the predominant frequency.  This  supports the  oblique pulsator model for this star as is the case with several other roAp stars. Probable evidence of the other rotational side-lobe and  the presence of other modes of pulsation are presented here. Details of observations and data analysis are given in the next section while results, discussion and conclusions are given in the remaining part of the paper.  

\section{Observations and data analysis}

The details of observations, data reductions and pulsation analysis are given in
the following subsections.

\begin{table}[t]
\caption{ Journal of high speed photometric observations of HD~12098. The
second column is the BJD at the center of the run, third column is the total
duration of observations. Last column lists the amplitude corresponding to the
frequency 2.1738~mHz for the individual nights (refer~sec. 2.3).}
\begin{center} 
\begin{tabular}{c|c|c|c} 
\hline 
Date         &Time(BJD)    & Duration&  Amp      \\ 
             &(2451000+)& (hours) & (mmag)    \\ \hline 
21 Nov 1999$^a$ & 504.27369   & 2.51    &  0.689    \\
20 Dec 1999$^a$ & 533.15988   & 2.00    & 0.381     \\
21 Dec 1999$^a$ & 534.18171   & 1.24    & 1.461     \\
22 Dec 1999$^a$ & 535.17755   & 3.12    & 1.412     \\
08 Oct 2000$^a$ & 826.30268   & 1.10    & 0.370     \\
11 Oct 2000$^a$ & 829.27440   & 2.64    & 1.199     \\
12 Oct 2000$^a$ & 830.25841   & 1.58    & 0.834     \\
15 Oct 2000$^a$ & 833.28734   & 2.01    & 0.673     \\
16 Oct 2000$^a$ & 834.94355   & 1.42    & 0.800     \\ 
18 Oct 2000$^b$ & 836.39091   &  5.19   & 1.054      \\
19 Oct 2000$^b$ & 837.33851   &  7.75   & 0.567      \\
20 Oct 2000$^b$ & 838.32158   &  7.76   & 0.534      \\
21 Oct 2000$^b$ & 839.31472   &  7.06   & 0.657      \\
22 Oct 2000$^b$ & 840.29415   &  7.82   & 1.904      \\
23 Oct 2000$^b$ & 841.31576   &  8.85   & 1.196      \\
07 Nov 2000$^a$ & 856.21624    & 2.64    & 1.469     \\ \cline{1-4}
\multicolumn{2}{c}{Total duration} & \multicolumn{1}{c}{64.99 hrs}& \\ \hline
\end{tabular}
\label{tab_jnl}
\end{center}
$^a${Observed at Naini~Tal}\\
$^b${Observed at Mt.~Abu}
\end{table}

\subsection{Data collection and Reduction}

The pulsations in the star HD~12098 were discovered on the night of 21/22
November 1999 during the roAp star survey observations at the 1.04~m Sampurnanand telescope of state observatory, Naini~Tal, India. Follow up observations were carried out at Naini~Tal and also at the 1.2~m telescope at Gurushikhar, Mt.~Abu, India. The log of observations is given in Table~\ref{tab_jnl}.

On both the telescopes we used a three channel photometer (Ashoka  et al.~\cite{ashoka01}). During the survey at Naini~Tal, we often use only the main
channel of the  photometer, while the follow up observations at Mt.~Abu are
conducted using all the  three channels,  monitoring a field star and sky
simultaneously. HD~12098 is a double star  with a companion which is 2.6 magnitude fainter and at an angular separation of $6''$ (Olsen~\cite{olsen80}). During observations,
both stars are kept   well within the diaphragm of $30''$.  The data are acquired as continuous 10~s integrations in  the Johnson B-filter using the  Quilt-9 software (Nather et al.~\cite{nather90}) running on a PC. Altogether, data for
HD~12098 are obtained on sixteen nights. Out of them, the nights from Naini~Tal
are with durations of observations ranging from 1.1 to 3.1 hours while the six
nights' data from Mt.~Abu are of relatively longer duration ranging from 7.06 to
8.85 hours except on one night where the duration is 5.19 hour. The data
obtained at both the places are corrected for dead time, sky  background and
earth's atmospheric extinction trend.

\begin{figure}[h]
\centering
\vskip -0.250in
\includegraphics[width=3.5in, height=3in]{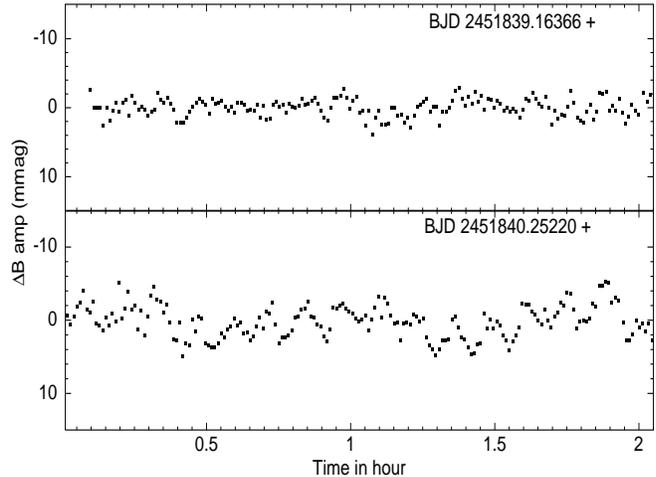}
\caption{ A representative  light curve of the star HD~12098 obtained on nights
BJD 2451839 and 2451840.  Each point represents the average of four 10~s
integrations in  the Johnson B filter.}
\label{fig_lc}
\end{figure}

The data sets are in time  expressed as barycentric Julian day (BJD) versus fractional intensity with respect to the mean intensity of the star.
Since the fractional  intensity variations are very
small these values are given as  mmag.  Figure~\ref{fig_lc}  presents  typical
light curve of the star HD~12098  obtained on 21/22 Oct 2000 and 22/23 Oct 2000.
The presence of the 7.6 minute oscillation and the day to day  variation in
pulsation amplitude is clearly evident in the light  curves.

\subsection{Pulsation analysis}

\begin{figure*}
\centering
\vskip -6.8in
\includegraphics[width=10.0in, height=10.50in]{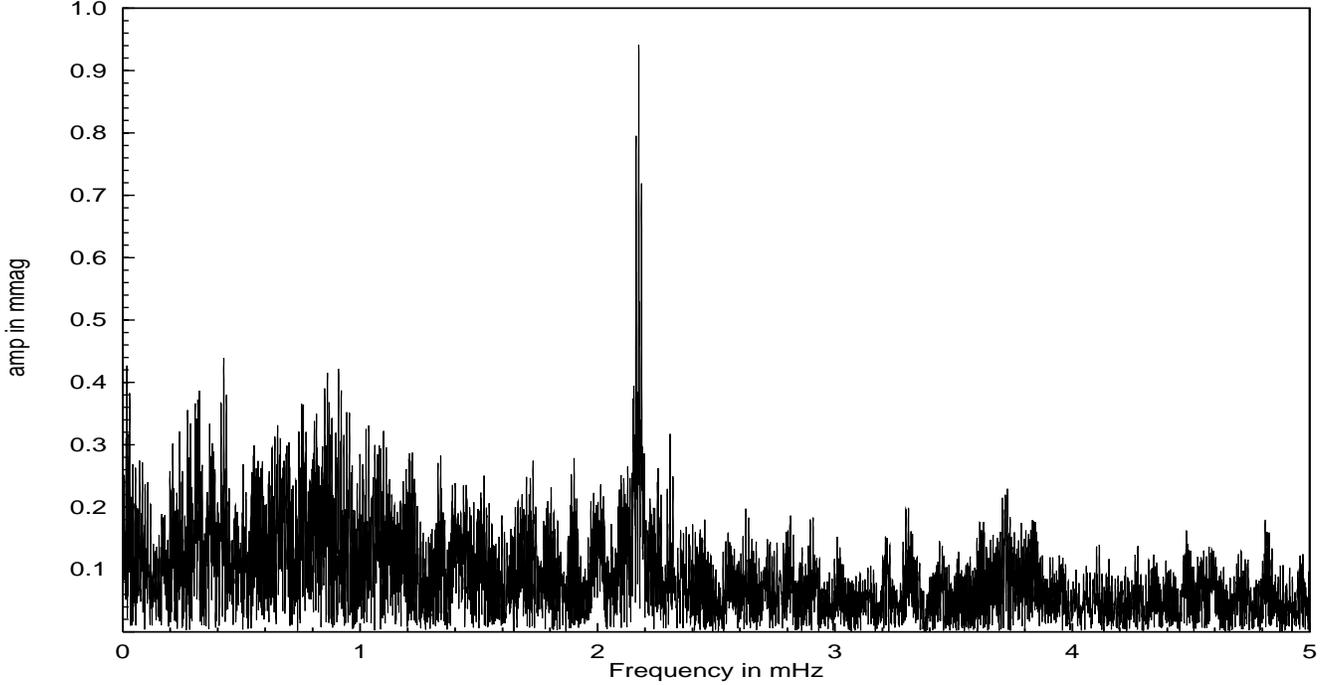}
\caption{The Fourier transform of the Mt.~Abu data set obtained during  BJD
2451836 - 841. It is evident  that the dominant frequencies located around 2.2~mHz are well above the noise level.} 
\label{fig_full} 
\end{figure*}

The presence of periodicity in the data-sets is searched for using the  Discrete
Fourier Transform (DFT) technique. All the individual runs exhibit a period
around 7.7 minutes with varying amplitudes. The data sets collected at Mt.~Abu,
all of which are of considerable length also exhibit the presence of a
frequency at 2.17~mHz in each of the runs, but with night to night variation in
amplitudes ranging from 0.5 to 1.9 mmag. This clearly indicates the likelihood
of the star being  multi-periodic. In order to resolve closely spaced
frequencies, all the runs from Mt.~Abu are concatenated together and subjected
to the DFT. Figure~\ref{fig_full} shows the amplitude spectrum of this data set
in the 0-5~mHz range. The dominant frequencies are around 2.2~mHz and there is
no excess power well above the noise level in any other region of the frequency
spectrum up to the Nyquist frequency (50~mHz).  There appear to be four frequencies present around 2.2~mHz which are given in column 1 and 2 of Table~\ref{tab_nlsq}.

\begin{figure*}
\centering
\vskip -0.50in
\includegraphics[height = 5.0in, width = 7.25in]{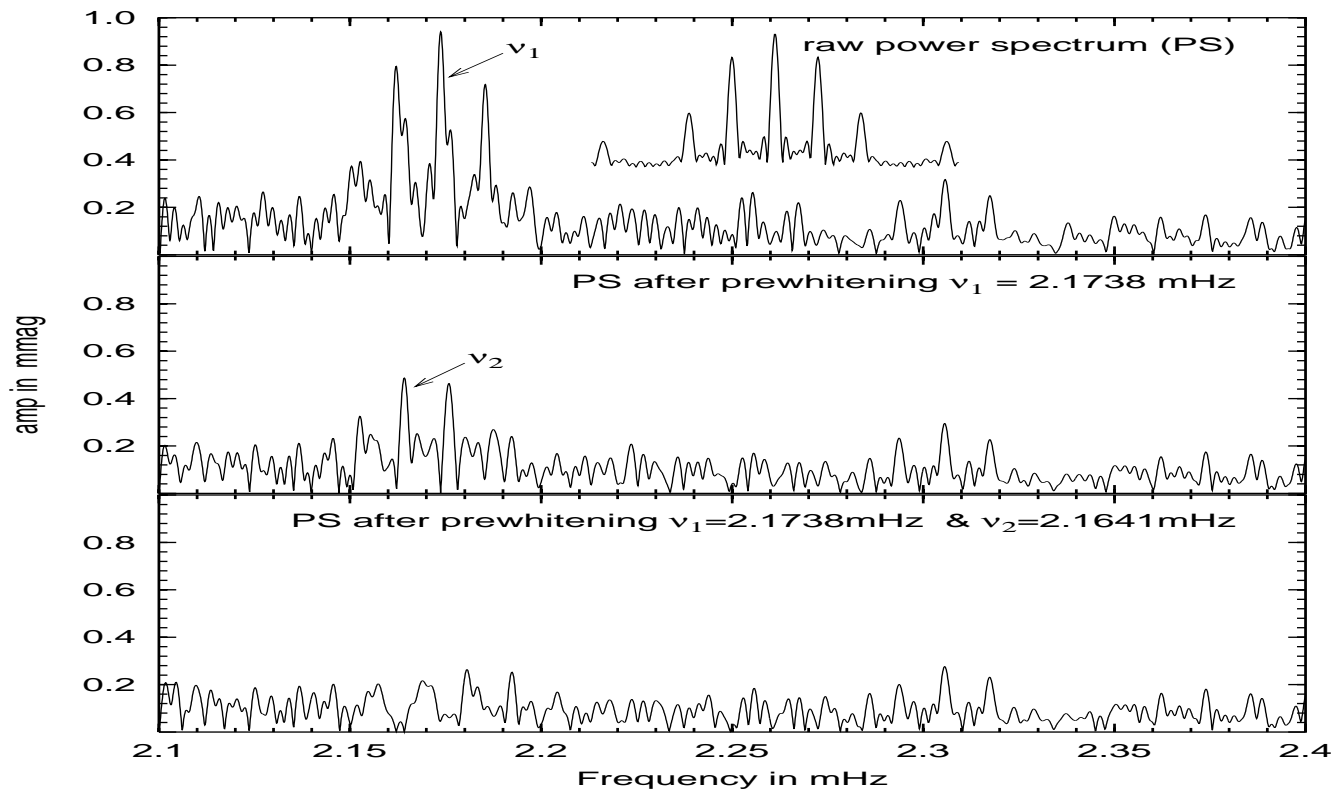}
\caption{A close-up view of the  FT of the data set BJD
2451836-841.  The second panel is the amplitude spectrum of the data  after
subtracting the frequency  $\nu_1 = 2.1738$~mHz from the time domain. The
third  panel is amplitude spectrum after subtracting the frequencies $\nu_1 =
2.1738$ and $\nu_2 = 2.1641$~mHz from the time domain.  The inset of panel~I
is the spectral window for the data-set calculated by sampling a noise-free
sinusoidal with the frequency and amplitude of the highest peak in
Fig.~\ref{fig_full}. } 
\label{fig_ft} 
\end{figure*}

\begin{table}[ht]
\caption{The dominant frequencies and amplitudes obtained for the Mt.~Abu data.
First two columns are the results of the  Fourier transforms (FT) and the  last
two columns are obtained by non-linear least square fit to the data.} 
\centering
\begin{tabular}{|c|c|c|c|}
\hline
\multicolumn{2}{|c|}{FT}         & \multicolumn{2}{c|}{NLSQ}         \\\hline
Frequency	& Amplitude	&Frequency	& Amplitude  	\\
 (mHz)		&  (mmag)	& (mHz)		&  (mmag)     \\\hline
2.17372	& 0.941	&2.17385$\pm 0.00006$  &0.918$\pm0.036$\\
2.16421	& 0.487	&2.16414$\pm 0.00010$  &0.511$\pm0.036$\\
2.18060	& 0.263	&2.18069$\pm 0.00016$  &0.276$\pm0.036$\\
2.30559	& 0.276	&2.30559$\pm 0.00014$  &0.279$\pm0.035$\\
 \hline
\end{tabular}
\label{tab_nlsq}
\end{table}

The  top panel of Fig.~\ref{fig_ft} shows an expanded view of the amplitude
spectrum around 2.2~mHz.  The  highest peak in this spectrum is at $\nu_1\, =\,
2.1737$~mHz with its 1 day alias pattern (the spectral window of the data set is
shown in the inset).   There is also an  indication of a second period at a
slightly lower frequency which is evident as  a double peak in $\nu_1$  and its
left side  alias.  In order to see whether it is an independent frequency or
not, we  prewhitened the main frequency $\nu_1$ from the time domain and
performed an FT again. The resultant amplitude spectrum of the residual data is
shown in the second panel of Fig.~\ref{fig_ft}. This clearly  establishes the
presence of a second frequency.
The peak with highest amplitude in this pattern corresponds
to a frequency $\nu_2 = 2.1641$~mHz. However, the amplitude of its alias at
2.1759~mHz is comparable. In FTs noise can alter the amplitudes of the
peaks. Hence we cannot unambiguously identify which of the above two is the real
frequency, although  the existence of a second frequency is evident.
Our identification of $\nu_2 = 2.1641$~mHz is thus subject to a 1/day alias
ambiguity.
 We again prewhitened this frequency $\nu_2$  from the data  set and the
amplitude spectrum  of the residual data is shown in the bottom panel of
Fig.~\ref{fig_ft}.  There is  an indication of two different frequency
groups in this final  spectrum, one around $\nu_3=2.181$~mHz and the other at
$\nu_4= 2.306$~mHz.  Both  of these frequencies are only marginally above the noise
level and are identified only by the typical window pattern they exhibit.
It may be noted however that while the spacing between $\nu_1$ and $\nu_2$ is about $9.5\, \mu$Hz,  the spacing between the central frequency $\nu_1$ and $\nu_3$ is around $7\, \mu$Hz. If 2.1759~mHz is identified with $\nu_2$ then the spacing between $ \nu_1 $ and $ \nu_2  $ is as small as 2.1 $\mu$Hz.  

The Mt.~Abu data set is subjected to a non-linear least square fit in
order to determine the errors on the frequencies and amplitudes of the
four frequencies determined from the FT.  The results of this analysis are
given in column 3 and 4 of Table~\ref{tab_nlsq}.

\subsection{Search for long term period}

The pulsation amplitudes of HD~12098 exhibit considerable variation from night
to night.  This is also indicative of long term periods. In order to search for
periodic modulation of the pulsation amplitude on a time scale of days, we
determine a nightly amplitude of the frequency $\nu_1$ by performing a linear
least square  (llsq) fit to each of the light curves acquired during 1999-2000.
The amplitudes thus determined are listed in the fourth column of
Table~\ref{tab_jnl}.

\begin{figure*}[t]
\centering
\includegraphics{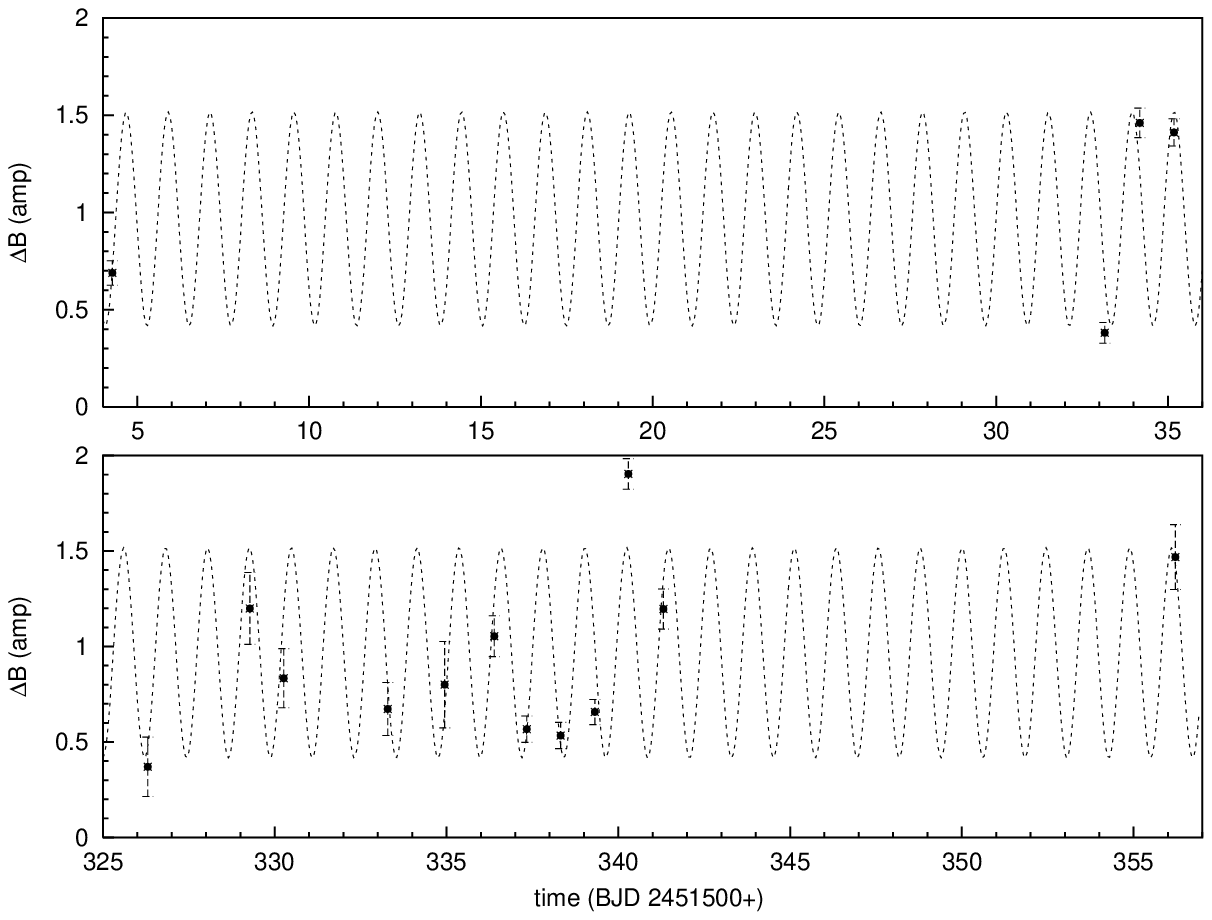}
\caption{Plot of pulsation amplitude of HD~12098 vs. BJD. X-axis  is the BJD at
the middle of the run and Y-axis is the pulsational amplitude in milli-magnitude. The points are the experimental results with the errors in the
amplitude. The dashed curve is the sinusoidal corresponding to the period 1.22
day. } 
\label{fig_mean} 
\end{figure*}

The resulting data-set of time  vs. amplitude (column 2 vs. column 4 of
Table~\ref{tab_jnl}) was  then subjected to a discrete Fourier transform,
non-linear least squares (nlsq) fit and a periodogram analysis using the
folding technique. 
All these methods yielded around five peaks, with the highest corresponding to a period of 1.22 day (0.82 cycles per day). 
This period though not completely convincing  provided best fit  (with a variance 17\% lower than the next best period) for the llsq fitted amplitudes from Table~\ref{tab_jnl} as shown in Fig.~\ref{fig_mean}.

\section{Results and Discussion}

The pulsation frequencies determined above are discussed below in terms  of
pulsation models.

\subsection{Frequency splitting and rotational period}

From the analysis of Mt.~Abu data, we have determined the presence of at  least
two dominant frequencies ($\nu_1=2.1738$ and $\nu_2= 2.1641$~mHz) in our data set along with two other frequencies ($\nu_3 = 2.1807$
and $\nu_4=2.3056$~mHz). Several roAp stars  are multi-periodic and
exhibit many modes (viz., Kurtz 1990). Hence the first hypothesis to test is
whether  or not these are individual modes.
The temperature  of this
star as estimated from both Str\"omgren indices and the spectral type, is
7820~$\pm$~30~K (Moon \& Dworetski~\cite{moon85}).  From the evolutionary models
the characteristic spacing  ($\Delta\nu$) between  alternate `$l$' and `$n$'
modes for Ap stars has been given by Heller \& Kawaler (\cite{heller88}) and
Shibahashi \& Saio (\cite{shibahashi85}) as a function of temperature. If the
two dominant frequencies in HD~12098 whose  temperature is 7820 K, have to be
independent modes  with a frequency spacing of only $9.5\, \mu$Hz or 
less, the theoretical models would  then imply a highly evolved star with a
luminosity in excess of $  2 \,\mathrm{L}_\odot$ and mass $ > 2.5\,\mathrm{M}_\odot$.
None of  the observational parameters show that HD~12098 is much more evolved than
the other  roAp stars. Moreover the typical spacing of alternate modes in
other roAp stars  ranges from $20\,\mu \mathrm{Hz}$ to $80 \,\mu \mathrm{Hz}$ (Kurtz, 1990). We
therefore conclude that the two dominant frequencies under discussion are
unlikely to be independent modes.

An  alternative explanation may be that, it is the `$m$' splitting due to
rotation in  $l=1$ state (Ledoux~\cite{ledoux51}) which gives rise to ($2l+1$) modes spaced in
frequencies as  
\begin{equation}
\nu_m - \nu_o = (1-C_{nl})m\Omega
\end{equation}
where $C_{nl}$ is a constant and $\Omega$  is the rotational frequency.
For HD~12098, we have no apriori estimate of the rotational period. If  we use
the theoretical values of $C_{nl}$ in the range  of  $0.01 \,\mathrm{to}\,0.001$
as given by Shibahashi \& Saio (1985) in Equ. (1), a $9.5\,\mu$Hz frequency
spacing implies a rotational period of around 1.21 to 1.22 day. In case the two
frequencies are caused due to `m' splitting, then the weak $2.1807$~mHz peak
could be the third of the expected triplet. Whether this hypothesis is valid for HD~12098, requires an independent determination of its rotation period. Incidentally, such a measurement has often led to this theory being disproved in the case of other roAp stars (Kurtz~\cite{kurtz90a}) favouring the oblique pulsator model instead.

 The pulsation  characteristics of  roAp stars are well understood in terms  of the oblique
pulsator model  (Kurtz~\cite{kurtz82}) and we apply it to HD~12098. The rotational
period expected from the $9.5\, \mu$Hz  frequency splitting of the two dominant
frequency peaks  is 1.22 day. In this case the third component of a 
conjectured l~=~1 frequency 
triplet can be expected  at $2.183$~mHz.  It may be noted that the unequal
frequency spacing of the two observed side-lobes with respect to the main peak may be due to the presence of magnetic field.  

Kurtz et al. (\cite{kurtz90b})  and Shibahashi \&
Takata (\cite{takata93}) have described a generalized oblique pulsator model in
which the effects of both the magnetic field and rotation are considered
(see also Dziembowski \& Goode (\cite{D&G85}, \cite{D&G86})).  The asymmetry in
the amplitudes of the triplet can be used to estimate the effect of magnetic
field vis-a-vis the rotational effect by the two parameters,

\begin{equation}
P_1 = \frac{A_{+1} + A_{-1}}{A_0} =  \tan{i}\ \tan{\beta}
\end{equation}
where $A_{-1}, A_0, A_{+1}$ are amplitudes of the triplet  corresponding to
frequency in increasing order,  $\beta$ is the  angle between the rotation and
magnetic axis (pulsation axis), $i$ is the inclination angle of the rotation
axis to the observer; and

\begin{equation}
P_2 = \frac{A_{+1} - A_{-1}}{A_{+1}+A_{-1}} =
\frac{C\Omega}{\omega_1^{(1)\mathrm{mag}}-\omega_0^{(1)\mathrm{mag}}} 
\end{equation}
where C is a constant, $\Omega$ is the rotation period and the  perturbations
due to the magnetic field depend on $|m|$ such that $\omega = \omega^{(0)} +
\omega_{|m|}^{(1)\mathrm{mag}}$. 
From the values given in Table~\ref{tab_nlsq}, we calculate $P_1 = 0.86$ and $P_2 = -0.29$.

If $\nu_1$ and $\nu_2$ are interpreted as components of a rotational multiplet in HD~12098, this would correspond to a rotation period of 1.22 day. This would then make HD~12098 the fastest rotating known roAp star.  Assuming a radius of $2R_\odot$ the  equatorial velocity   is calculated to  be $\approx$ 80 km s$^{-1}$.  For Ap stars the limit of $v\sin i$ is of the order of 10 km s$^{-1}$ (Wolff~\cite{sidney83}). Hence for HD~12098, this would imply an $i \approx 10^\circ$ and  $\beta\, \approx \,78^\circ$. Based on the magnetic
field studies, observations of pole reversal would indicate $\beta \ge
80 ^\circ$. However, we remind the reader that $\nu_2$ is subject to a 1
cycle/day alias ambiguity.
In the event that $ \nu_2 $ is identified with 2.1759~mHz instead of 2.1641~mHz,
the spacing between  $ \nu_1 $ and $ \nu_2 $ would be 2.1 $ \mu $Hz. This would
imply a rotational period of 5.51 day.
Our present data cannot distinguish between these possibilities, and the
issue will only be resolved by acquiring more data.

\subsection{Other modes}

  The other frequency at the limit of detection in our complete data set
  is at $2.3056$~mHz.  This peak also exhibits the typical alias pattern
  and is comparable in amplitude to the $2.1807$~mHz peak.  It may be noted however
  that this frequency has been detected in some of our earlier data-sets.
  The runs in which this frequency is predominantly above the noise level
  corresponds to observations on 
 BJD 2451504.27 and  533.16.  Hence this
  may be either a short lived mode or a mode which is severely modulated.
  In addition we have also detected excess power around 4~mHz on
  several nights.

\subsection{HD~12098 and  other roAp stars.}

HD~12098 appears to exhibit a simple  power spectrum at first glance. There
are several roAp stars such as HD~196470 and  HD~193756 (Martinez et al.~\cite{peter91}) which exhibit only a single frequency. However, 
HD~12098 appears to have  few modes exhibiting effects of  rotation in the
star. Since the amplitude of the main pulsation does not exceed 2 mmag, it is
one of the low amplitude roAp pulsator. Further observations may yield other low
amplitude modes in the star,  if there are any. If the rotation period of the
star is 1.22 day or close to it,  then it becomes the fastest rotating roAp
star.   HD~6532 which has a rotational period of 1.9455 day (Kurtz \& Marang~
\cite{kurtz87a}) also has a similar amplitude spectrum with six frequency peaks
(Kurtz \& Cropper \cite{kurtz87b}). Another roAp star which is very similar is
HD~80316 (Kurtz \cite{kurtz90b}). The rotation period of this star is not yet
confirmed, but is probably around 2 days.

The estimated  $\delta m_1$ (= -0.122) for HD~12098 
is at the extreme edge of the selection criteria which sets the limits
for probable roAp stars for this index (Martinez \& Kurtz 1995).  However there
are other roAp stars with even more negative $\delta m_1$ (e.g., HD~101065). In
this respect  it may be noted that HD~80316 has $\delta m_1 =
-0.118$. The principal oscillation frequency of this star is 2.2516~mHz with an
amplitude of 0.44 mmag.  Hence HD~80316 can be considered very similar to
HD~12098 in several observed properties including the frequency spectrum and the
pulsation amplitude.

\section{Conclusions}

We have presented the photometric observations of  a new roAp star
 HD~12098 discovered in the ``Naini~Tal-Cape survey for northern hemisphere roAp
stars''. The detailed analysis of the observations shows the presence of
two predominant frequencies. The main frequency is at $2.1738$~mHz
corresponding to a period of 7.67 minutes  with an amplitude of 0.918 mmag.
The second frequency can be either 2.1641 or 2.1759~mHz. This is most likely 
the rotational side-lobe of the main frequency $\nu_1$.

Hence we propose that HD~12098 is an oblique pulsator which pulsates in the $l=1$ mode with a main period of 7.67 minute. The rotation period of this star
is of the order of a few days, but with our data we are unable to discriminate
between possible rotation periods of 1.2~d or 5.5~d. Independent
determination of the rotational period using differential photometry, or
additional multi-site high-speed observations are required to resolve this
ambiguity and to improve our understanding of the pulsation spectrum of HD~12098.

\section*{Acknowledgments}

The authors gratefully acknowledge the support of the  observing and
engineering staff at PRL Mt.~Abu observatory and State Observatory,
Naini~Tal. PM and DWK acknowledge the gracious hospitality and support of State
Observatory, Naini Tal, and the ISRO Satellite Centre, Bangalore. Thanks are
also due to Dr. M.M. Dworetski for providing the programs to estimate the
effective temperature of the star. PM acknowledges financial support from
the South African National Research Foundation for travel to India.

\end{document}